\documentclass[aps,twocolumn,showpacs,superscriptaddress,floatfix,letter]{revtex4-1}
\usepackage{graphicx}
\usepackage{psfrag}
\usepackage{graphicx,amssymb,amsmath}
\begin{document}
\newcommand{\beq}{\begin{equation}}
\newcommand{\eeq}{\end{equation}}
\newcommand{\ben}{\begin{eqnarray}}
\newcommand{\een}{\end{eqnarray}}
\newcommand{\bea}{\begin{array}}
\newcommand{\eea}{\end{array}}
\newcommand{\om}{(\omega )}
\newcommand{\bef}{\begin{figure}}
\newcommand{\eef}{\end{figure}}
\newcommand{\leg}[1]{\caption{\protect\rm{\protect\footnotesize{#1}}}}
\newcommand{\ew}[1]{\langle{#1}\rangle}
\newcommand{\be}[1]{\mid\!{#1}\!\mid}
\newcommand{\no}{\nonumber}
\newcommand{\etal}{{\em et~al }}
\newcommand{\geff}{g_{\mbox{\it{\scriptsize{eff}}}}}
\newcommand{\da}[1]{{#1}^\dagger}
\newcommand{\cf}{{\it cf.\/}\ }
\newcommand{\ie}{{\it i.e.\/}\ }   

\newcommand{\spazio}{\vspace{0.3cm}}
\hyphenation{bio-mol-ecules}
\newcommand{\de}[1]{\frac{\partial}{\partial{#1}}}
\newcommand{\U}{\tilde{U}}
\newcommand{\V}{\tilde{V}}

\title{Focusing in Multiwell Potentials: Applications to Ion Channels}

\author{L.~Ponzoni}
\affiliation{Dipartimento di Matematica e
Fisica, Universit\`a Cattolica, via Musei 41, 25121 Brescia, Italy}
\author{G.~L.~Celardo}
\affiliation{Dipartimento di Matematica e
Fisica and Interdisciplinary Laboratories for Advanced Materials Physics,
 Universit\`a Cattolica, via Musei 41, 25121 Brescia, Italy
and I.N.F.N., Sezione di Pavia, Italy}
\author{F.~Borgonovi}
\affiliation{Dipartimento di Matematica e
Fisica and Interdisciplinary Laboratories for Advanced Materials Physics,
 Universit\`a Cattolica, via Musei 41, 25121 Brescia, Italy
and I.N.F.N., Sezione di Pavia, Italy}
\author{L.~Kaplan}
\affiliation{Tulane University, Department of Physics, New Orleans, Louisiana 70118, USA}
\author{A.~Kargol}
\affiliation{Loyola University, Department of Physics, New Orleans, Louisiana 70118, USA}

 \begin{abstract}       
 We investigate out of equilibrium stationary distributions induced by a stochastic dichotomous noise on double and multi-well models for ion channels.  
 Ion-channel dynamics is analyzed both through over-damped Langevin equations and master equations. 
As a consequence of the external stochastic noise, we prove a non trivial focusing effect,
 namely the probability distribution is concentrated only on one state of the multi-well model.
We also show that this focusing effect , which occurs at physiological conditions, cannot be predicted by a simple master equation approach. 
\end{abstract}                                                               
                                                                            
\date{\today}
\maketitle

 \section{Introduction}

Interaction of physical, chemical or biological systems with fluctuating stimuli, both random and/or periodic, has in recent years received a lot of attention in a variety of contexts. It gives rise to new effects in nonlinear systems that cannot be observed under stationary perturbation. 
One of the most studied examples of such phenomena is the stochastic resonance \cite{Gammaitoni, kallunki, bezrukov, hanggi, adair, goychuk1, goychuk2, wiesenfeld}. It is an effect when a noise with suitable properties, either intrinsic or added to the system, improves the system response to weak time-dependent signals. The signal-to-noise ratio can paradoxically be boosted by increasing the level of noise present in the system. The SR was first suggested as an explanation for the cyclic ice ages observed in the earth’s climate \cite{Gammaitoni} but has been since expanded as a plausible explanation of various effects in other areas. Of particular interest seems to be the application to explain the detection of very weak signals in noisy environments observed in biological systems, e.g. ion channels \cite{bezrukov, adair, goychuk1}.

Another example is the dynamics of ratchets. The term refers to spatially asymmetric potentials which, interacting with Brownian fluctuations generate a directed drift in the system. The scale of the effect can be controlled by adjusting the properties of the noise. A similar ratchet-like effect has been observed in spatially symmetric potentials subject to temporally asymmetric noise (so-called correlation ratchets) \cite{chacron}.

An interesting example of such fluctuation-induced phenomena is called the resonant activation \cite{doering, flomenbom}. It is an effect where the diffusion over a fluctuating potential barrier is correlated with the fluctuation rate and a resonant effect is observed. As noted in \cite{doering} if the barrier fluctuations are slow compared to the natural time scale of barrier crossing then the effect can be described using an adiabatic approximation. For very fast oscillations the barrier crossing rate would be that for the average barrier height. For intermediate barrier fluctuation rates a resonance-like enhancement of barrier crossing rate is observed. This has been applied e.g. to chemical reaction rates \cite{schmid} and ion transport through channels \cite{lee}.

A related effect, and the one we investigate in this paper is the nonequilibrium kinetic focusing first proposed in \cite{Millonas}. It is an effect where a particle moving in a ratchet-like potential subject to external dichotomous noise fluctuations gets trapped in one of the energy wells. This mechanism was applied to gating kinetics of voltage-gated ion channels, achieving a focusing of the channels into a particular conformational state corresponding to this energy well. First experimental study 
of this effect in Shaker K+ channels was reported in \cite{Armin}. The phenomenon of particle trapping in correlation ratchet was also investigated in \cite{chacron}.

Ion channels are membrane proteins in biological cells that form gated channels for a controlled exchange of physiologically important ions, such as sodium or potassium, between the cytosol and the extracellular medium \cite{hille}. They play a very important role in various physiological processes (nerve impulses, muscle contraction etc.), and several human and animal disorders have been linked to malfunctioning ion channels. Therefore there have been numerous studies aimed at better understanding of channel gating kinetics and ultimately – controlling it. Kinetic focusing of ion channels seems to be very promising for both of these goals.

Gating kinetics of voltage-gated ion channels is a reflection of conformational changes of the channel molecule. From a physico-chemical standpoint the gating process has been modeled as a particle (so-called gating particle) moving in a certain multi-well energy landscape (the energy of molecular conformational states). Mathematically, the most commonly used description is a discrete Markov chain, where different Markov states correspond to minima in the energy profile and the transition rates between the states reflect the height and width of the energy barrier. The time evolution of the system is described by the master equation. These types of models dominate the literature about ion channels and have been reasonably successful in explaining physiologically relevant channel effects (see e.g. \cite{Bezanilla}). However, they are coarse approximations only, completely disregard intra-well motions of the gating particle, and work only if the energy barriers separating the wells are sufficiently high. More accurately the process can be described by the overdamped Langevin equation \cite{Gammaitoni}.

In this paper we study both approaches to simple models of ion channel gating. In Section $II$ we introduce the model,  in Section $III$ we consider a two-state (two-well) model which may represent an ion channel with two stable conformational states only:  one open (where the ions can go through the cell membrane) and one closed. 
 In Section $IV$ we analyze a more realistic model of gating kinetics for a {\it Shaker} potassium channel developed by Bezanilla et al. \cite{Bezanilla} based on patch-clamping experiments with ionic and gating currents. This model was also used in \cite{Millonas} to show the existence of the focusing effect.

%
%
%
%
%

%

\section{The Model}


In the model we use, the  potential minima are the possible states of the ion channel,
 while the barriers heights between two neighboring states are
 chosen so to reproduce the correct transition rates. 
 It is  is described by  the overdamped Langevin equation
\begin{equation}
 \gamma\dot{x}(t) = -u'(x) + \sqrt{2\gamma k_BT}\xi(t)
\end{equation}
where $\xi(t)$  is a white noise with $\langle \xi(t) \rangle=0$ and 
$\langle \xi(t) \xi(0) \rangle=\delta(t)$, $u(x)$ is the potential landscape, $\gamma$ is the
friction coefficient and $T$ is the temperature of the bath.

We will consider the effect of a time dependent stochastic external potential on the 
out of equilibrium stationary distributions of our model. 
When an external voltage, $v_{ext}(t)$, is applied across the cell  membrane, the particle behaves as if 
it had a charge valence, $z$, and the Langevin equation becomes (see \cite{Millonas}),
\begin{equation}
 \gamma\dot{x}(t) = -u'(x) + zv_{ext}(t)/\lambda  + \sqrt{2\gamma k_BT}\xi(t),
\end{equation}
where the length scale, $\lambda$, is the cell membrane thickness. Using the rescaled variables
\begin{equation}\label{rescaling}
\begin{split}
& X=\frac{zx}{\lambda}, \quad \tau=\frac{z^2k_BT}{\gamma \lambda^2}t, \\
& V_{ext}(t)=\frac{v_{ext}(t)}{k_BT}, \quad U(x)=\frac{u(x)}{k_BT},
\end{split} \end{equation}
we finally get the Langevin equation in the dimensionless form
\begin{equation} \label{LangMill}
 \dot{X}(\tau) = -U'(X) + V_{ext}(\tau) + \sqrt{2}\,\xi(\tau).
\end{equation}
In this equation we see that the membrane potential, $V_{ext}$, changes the potential shape in a linear way:
\begin{equation}
U(X) \rightarrow U(X)-V_{ext} \, X.
\end{equation}

The specific choice of the external stochastic potential is a 
a  dichotomous noise : 
\begin{equation}
V_{ext}(\tau)=V_0 + V_{DN}(\tau),
\label{Vext}
\end{equation}
where $V_0$ is a constant term and $V_{DN}(\tau)$ represents a is time-dependent 
random and asymmetric switching between to fixed values of the potential, $V_\pm$,:

 \begin{equation}
  V_+ = \sqrt{\frac{D}{\tau_v}\left(\frac{1+\epsilon}{1-\epsilon}\right)}, \qquad 
 V_- = -\sqrt{\frac{D}{\tau_v}\left(\frac{1-\epsilon}{1+\epsilon}\right)}
 \end{equation}
with transition probabilities,
$w_+$ (from plus to minus state), and $w_-$ (from minus to plus state),
 \begin{equation}
  w_+ = (1+\epsilon)/2 \tau_v,	\qquad	w_- = (1-\epsilon)/2 \tau_v.
  \label{rates}
 \end{equation}
It is easy to show that such  DN 
 has a null time average
\begin{equation}
\langle V_{DN}(\tau) \rangle = 0,
\label{aveV}
\end{equation}
and correlation function 
$$\langle V_{DN}(\tau) \, V_{DN}(0) \rangle = (D/\tau_v)\exp(-\tau/\tau_v),$$ 
so that it can be fully characterized by three  dimensionless parameters: 
\textit{correlation time} $\tau_v$,  \textit{intensity} or \textit{amplitude } $D/\tau_v$ 
and  \textit{asymmetry} $\epsilon$ (with $-1<\epsilon <1$).

\section{Focusing in a double-well potential}

In this Section we consider a  two state model for an ion channel, represeting open/close configurations.
Even if it is too simple to adequately represent the function of real channels,
It is commonly used as  toy models,  to demonstrate various phenomena \cite{mahato1, schmid, ghosh, dicera, mahato2}. 
A two-state model can also be understood as a sub-model of a larger system, see Section $IV$. 

\begin{figure}[b!]
\spazio
\centering
\includegraphics[width=0.5 \textwidth]{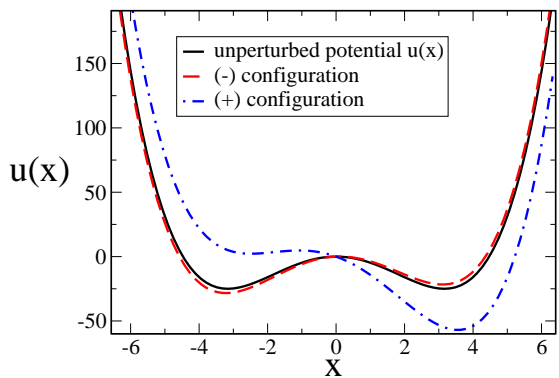}
\caption{(Color online) 
The unperturbed double-well potential $u(x)$ of Eq.~(\ref{eq1}) 
and the two configurations of the potential when it is modified by the 
dichotomous noise, as indicated in the legend.
Potential parameters are $a=10$ and $b=1$, while the DN shown refers to: $D/\tau_v=0.71 $, $k_BT=0.15 \Delta u$, $\epsilon=0.8$.}
\label{potential}
\end{figure}

The  double well potential landscape here considered is: 
\begin{equation} \label{eq1}
 u(x) = -\frac{1}{2}ax^2 + \frac{1}{4}bx^4,
\end{equation}
which has two minima in $\pm x_m = \pm \sqrt{a/b}$, and a central barrier 
with height $\Delta u = a^2/(4b)$; see Fig.~\ref{potential}. 

Due to thermal fluctuations, the particle is forced to jump between the two wells with a rate that for $\Delta u/k_BT \gg 1$, is given by the \textit{Kramers formula} \cite{Gammaitoni}:
\begin{equation} \label{Kramers}
 w_K = \frac{\omega_0 \omega_b}{2\pi\gamma} \exp \left(-\frac{\Delta u}{k_BT}\right),
\end{equation}
where $\omega_0^2 = |u''(x_m)|$, $\omega_b^2 = |u''(x_b)|$ (evaluated at the top $x_b$ of the central barrier).

In this Section we consider the effect of a DN perturbation with $V_0=0$, see Eq.(\ref{Vext}). 
Depending on the state of the DN, there are  two possible configurations for the potential: 
a $(-)$ configuration and a $(+)$ configuration, as depicted in Fig.~\ref{potential}. 
For each configuration there are two different thermal transition rates: for example, 
in the $(-)$ configuration  $w^L(-)$ stands  for the transitions from the left to the right
 well and $w^R(-)$ for the transitions from the right to the left well; and similarly in the $(+)$ configuration.

In order to study the stationary probability distribution one has to solve the stochastic equation, Eq.(\ref{LangMill}).
In our case the Langevin equation  (\ref{LangMill})  has been solved numerically with 
an \textit{order 1.5 strong Ito-Taylor scheme} (in explicit form), 
described by Platen and Wagner in \cite{Platen}.
Starting with different initial conditions (random distribution of initial position) and integrating in time the Langevin equation,
 it is possible to compute the probability
$P_{L/R}(\tau)$ to be found at the time $\tau$ in each potential well.
After a transient time, we verified that this probability reaches a stationary  value, 
$P^{eq}_L$,   dependent on the DN parameters.

\begin{figure}[h!]
\spazio
\centering
\includegraphics[width=0.54 \textwidth]{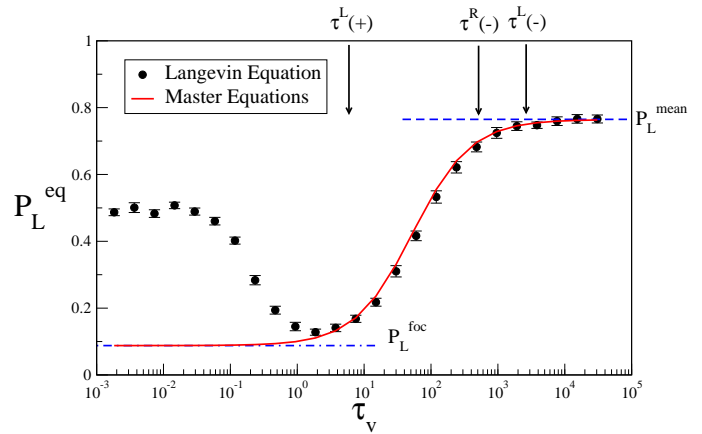}
\caption{(Color online) Equilibrium probability in the left well versus the DN correlation time $\tau_v$. 
Full (black) circles have been obtained from the Langevin equation, 
while full (red) line has been obtained from a master equation approach.
Thermal transition times $\tau^{L,R}(\pm)$ are shown as arrows
(except from
$\tau^R(+)$, which is found to be of order $10^{6}$ and is out of scale).
The  horizontal upper dashed line represents the limit $P_L^{mean}$, computed with Eq.~(\ref{limitedx}),
while the dot-dashed horizontal lower line is$P_L^{foc}$, see Eq.(\ref{PLfoc}).
Here we have $\epsilon=0.8$, $k_BT = 0.15 \Delta u$,  $D/\tau_v=0.71$,
as in  Fig.~\ref{potential}.}
\label{minimo}
\end{figure}

Because of the DN, the potential switches between two configurations, 
so it is possible to define  
four thermal transition times in the system: 
 $\tau^{L/R}(+/-)$, where, for example, $\tau^L(+)$ is the mean time after which a particle jumps from the left to the right well when the potential is in the $(+)$ configuration (that is when $V_{DN}=V_+$). 
Note that in this Section we consider  a regime  in which both the DN amplitude and the thermal fluctuations are  not too large,
compared to the height of the potential barrier.
Small thermal fluctuations w.r.t. the potential barrier correspond to physiological conditions at room temperature for
ion channels, \cite{Millonas}. Moreover also moderate DN amplitude is a realistic regime since 
 a large amplitude would destroy the ion-channels. 
Under these assumptions, 
 the thermal transition times can be computed through the 
Kramers rates,  $w^{R/L}(\pm)$ see Eq.(\ref{Kramers}),  for the perturbed potential,   so that we have: 
\begin{equation}
 \tau^{L/R}(\pm)=1/w^{L/R}(\pm).
\end{equation}
We should also take into account that 
there are two more  characteristic times, given by the DN, namely,
\begin{equation}
 \tau_v^{\pm} = \frac{1}{w_{\pm}} = \frac{2}{(1{\pm} \epsilon)} \tau_v,	
\end{equation}
where $w_{\pm}$ are given by Eq.(\ref{rates}). 
Note that $ \tau_v^{\pm}$  are an estimate of the times spent in each configuration between two DN switchings.

The results of Langevin equation   are shown  in Fig.~\ref{minimo}  at 
fixed  DN intensity, $D/\tau_v$,   asymmetry, $\epsilon$,
and varying the DN correlation time, $\tau_v$.
Each data in the graph represents the left well equilibrium probability, $P_L^{eq}$.
As one can see there is a minimum of  $P_L^{eq}$, which means that the $90\%$ of probability is
concentrate in the right well, at some peculiar value of $\tau_v$.
This is the  \textit{nonequilibrium kinetic focusing}, which  should not be confused with the trivial 
focusing which occurs at large $\tau_v$, where $P_L^{eq} \sim 0.8$. Indeed this
focusing for large $\tau_v$ can be understood based on equilibrium consideration.
Indeed, when the DN
transition times, $\tau_v^\pm$, are 
much larger than the thermal transition times, $\tau^{L,R}$,
we can assume that the system
reaches separately the thermal equilibrium  in each potential configuration.

Therefore if $P_{L/R}(\pm)$ are the thermal Gibbs  probabilities for each well in the $(+)$ and $(-)$ 
configuration, then it is reasonable to assume that,
for $\tau_v^{\pm} \rightarrow \infty$, the probability $P_L^{eq}$ approaches the mean value 
\begin{equation} \label{limitedx}
P_L^{eq} \rightarrow \frac{\tau_v^+}{(\tau_v^+ + \tau_v^-)}P_L
(+)+\frac{\tau_v^-}{(\tau_v^+ + \tau_v^-)}P_L(-)  \equiv P_L^{mean}.
\end{equation}
$P_L^{mean}$ is shown as the dashed horizontal line in Fig.~\ref{minimo} and agrees very well with
numerical data.

On the other side, in the limit of fast switching of the DN ($\tau_v \rightarrow 0$), the probability distribution becomes symmetric:
\begin{equation}\label{limitesx}
 P_L^{eq} , P_R^{eq} \rightarrow 1/2,
\end{equation}
which is the value obtained with our numerical simulations of the Langevin equation,
see  Fig.~\ref{minimo}.
The limit  $\tau_v \rightarrow 0$ can be understood 
 using the Langevin equation, by assuming, for simplicity, a periodic, rather than a stochastic  switching,
 between the two potential configurations.
Indeed,
after the first step, assuming we are in the $(+)$ configuration, we can write, see Eq.(\ref{LangMill}):
\begin{equation}
 (\Delta X)^+ = [-U'(X) + V_+   + \sqrt{2  } \, \xi] \tau_v^+,
\end{equation}
and after the second step
\begin{equation}
 (\Delta X)^- = [-U'(X) + V_-   + \sqrt{2  } \, \xi] \tau_v^-,
\end{equation}
neglecting the terms proportional to $\tau_v^+\tau_v^-$. The total variation after a time $\tau_v^+ + \tau_v^-$ is then:
\begin{equation}\begin{split}
 \Delta X & = (\Delta X)^+ + (\Delta X)^- \\
 & = [-U'(x) + \sqrt{2 } \, \xi] (\tau_v^+ + \tau_v^-) + (v_+\tau_v^+ + v_-\tau_v^-).
 \end{split}
\end{equation}
and, dividing by $\tau_v^+ + \tau_v^-$, we get the Langevin equation for a fast oscillating potential:
\begin{equation}
 \dot{X} = -U'(X) + \sqrt{2 } \, \xi + \frac{V^+\tau_v^+ + V^-\tau_v^-}{\tau_v^+ + \tau_v^-}.
\end{equation}
The last term corresponds to the time average of the DN, which is zero (see Eq. (\ref{aveV})): 
this proves that for fast oscillations of the perturbing potential the 
Langevin equation does not feel the perturbation, namely it behaves symmetrically,
as anticipated.

In the intermediate  regime we have a non trivial result:
from  Fig.~\ref{minimo}, it is clear that
 the left well probability $P_L^{eq}$ has a minimum 
(i.e. there is a ``focusing" in the right well) for a particular choice of the DN correlation time $\tau_v$. 
The focusing effect is an interesting result, since the
applied DN has zero mean (Eq.(\ref{aveV}), and nevertheless it can induce stationary
distributions which focus the system in one well.

In order to understand what are the conditions for the focusing,
we can attempt a simple explanation through a master equation approach:
\begin{equation}\label{mesystem}
 \dot{\mathbf{P}} = W(\tau) \, \mathbf{P},
\end{equation}
where $\mathbf{P}(\tau)=(P_L(\tau), P_R(\tau))$  and the time dependent transition matrix $W(\tau)$, which switches between the two configurations :
\begin{equation}
 W(\pm)= \left(
\begin{array}{cc}
 -w^L(\pm) & w^R(\pm) \\
 w^L(\pm) & -w^R(\pm) \\
\end{array}
\right)
\end{equation}
with the same rates as in Eq.~(\ref{rates}).
In Fig.~\ref{minimo} we   compare the  results obtained through the Langevin equation with those
obtained via  the master equation, Eq.(\ref{mesystem}).
As one can  see,  there is a good agreement between the two approaches
when the  DN correlation
 time $\tau_v$ is larger  than the value at  which the best focusing occurs, while,
when $\tau_v \rightarrow 0$ the master 
equation gives a complete different result.

This is a limitation of the master equation formalism, 
for which we can give an explanation by assuming, as was done above for the Langevin equation, a periodic switching.
At the initial time the matrix is in the $W(+)$ configuration, and after a time $\tau_v^+$ it changes into the $W(-)$ configuration, where it stays for a time $\tau_v^-$; and so on. Thus assuming that $\tau_v^+$ and $\tau_v^-$ are both infinitesimal, after the first step we can write:
\begin{equation}
 (\Delta \mathbf{P})^+ = W(+) \, \tau_v^+ \, \mathbf{P}(0),
\end{equation}
where $(\Delta \mathbf{P})^+ = \mathbf{P}(\tau_v^+)-\mathbf{P}(0)$. Similarly, after the second step we have:
\begin{equation}
  (\Delta \mathbf{P})^- = W(-) \, \tau_v^- \, \mathbf{P}(\tau_v^+) = W(-) \, \tau_v^- \, \mathbf{P}(0),
\end{equation}
where the last term holds neglecting the terms proportional to $\tau_v^+\tau_v^-$. The 
total variation after a time $\tau_v^+ + \tau_v^-$ is:
\begin{equation}\begin{split}
\Delta \mathbf{P} & =  (\Delta \mathbf{P})^+ + (\Delta \mathbf{P})^- \\
& = [W(+)\tau_v^+  + W(-) \tau_v^-]\, \mathbf{P}(0),
\end{split}
\end{equation}
and, dividing both sides by $\tau_v^+ + \tau_v^-$, we get a system of differential equations with an \textit{averaged} transition matrix:
\begin{equation} \label{averagedmatrix}
 \langle W \rangle = \frac{W(+)\tau_v^+  + W(-) \tau_v^-}{\tau_v^+ + \tau_v^-}.
\end{equation}
This averaged matrix is different from the unperturbed 
transition matrix, that  we would expect for $\tau_v \rightarrow 0$, and determines the following equation for the 
probabilities: 
\begin{equation}\begin{split}
 \dot{P_L} & = \langle W \rangle_{11} \, P_L + \langle W \rangle_{12} \, P_R \\
           & = (\langle W \rangle_{11} - \langle W \rangle_{12}) P_L + \langle W \rangle_{12}.
\end{split}
\end{equation}
and the stationary solutions $\dot{P}_L^{eq}=0$, is given by:
\begin{equation} 
\label{PLfoc}
 P_L^{eq} = \frac{\langle W \rangle_{12}}{\langle W \rangle_{12}-\langle W \rangle_{11}} \equiv P_L^{foc}.
\end{equation}
Therefore this average transition matrix in general does not have  a symmetric solution.
In  Fig.~\ref{minimo} the value of  $P_L^{foc}$ is shown to agree with the master equation solution in the limit $\tau_v \rightarrow 0$,
and, interestingly, it is close to the focusing value of  $P_L^{eq}$ at the minimum.  
This coincidence will be used below to determine the dependence of the focusing intensity
on the parameters of the model.

An intuitive explanation of the failure of  the master equation approach for  $\tau_v \rightarrow 0$, is that  the 
system does not have time to equilibrate in each well before the switching and thus the two-state model
becomes inappropriate.

The fact that a master equation approach  cannot explain the focusing effect which occurs
for fast switching of the DN, shows that the focusing is highly non trivial, being a truly non-equilibrium effect. 
For a better understand this focusing we need to consider 
the interplay of several time scales.
Physically, we can expect focusing when $\tau_v^+$ 
(mean time for which the potential is in $(+)$ configuration) is of the same order than 
$\tau^L(+)$ (i.e. the time the particles need to jump towards the right 
well in the $(+)$ configuration)
and, at the same time,   $\tau_v^-$ should be  much smaller than the time $\tau^R(-)$ 
the particles need to jump towards the left well. Roughly speaking we  can thus
write
the \textit{focusing conditions} at the right well as,
\begin{equation} \label{focusingcond}
\left\{ \begin{aligned}
\tau_v^+   \approx \tau^L(+) \ll \tau^R(+) \\
\tau_v^- \ll  \tau^R(-) \ll \tau^L(-).
\end{aligned} \right.
\end{equation}


\begin{figure}
 \centering
 \includegraphics[width=0.54 \textwidth]{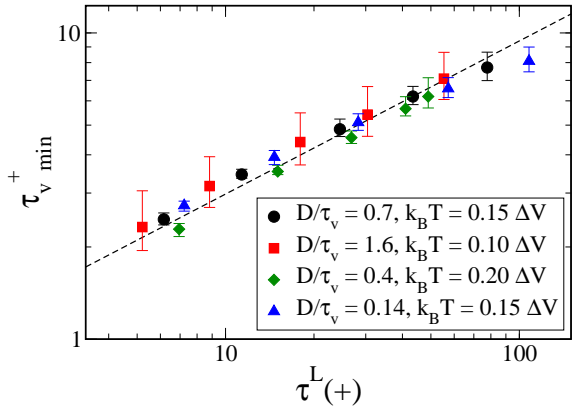}
  \caption{(Color online) Correlation time ${\tau_v^+}_{min}$, 
for which we observe focusing, {\it vs }
 the transition time $\tau^L(+)$. Each data set 
refers to simulations with different $\epsilon$ 
but  same intensity $D/\tau_v$ and temperature $T$ as
indicated in the legend. The dashed line indicates a square root dependence.
Error bars describe the variation of ${\tau_v^+}_{min}$
for which $P_L^{eq}$ changes by  0.1\% from its minimum value.
}
\label{log}
\end{figure}

We now study  
how  the switching time,  ${\tau_v^+}_{min}$, at which a minimum
 of $P_L^{eq}$ is found, depends 
 on the thermal transition time, $\tau^L(+)$ for different 
asymmetry  $\epsilon$, the intensity $D/\tau_v$ 
and  the temperature $T$.
Data are shown in Fig.~\ref{log}. 
Each value of ${\tau_v^+}_{min}$ has been 
computed from a graph like Fig.~\ref{minimo}, by making a 
quadratic fit of the function $P_L^{eq}(\log(\tau_v))$ around its minimum.
In the same figure each curve at fixed  $D/\tau_v$ and $k_B T$, has been obtained 
varying the asymmetry parameter $\epsilon$. 
We have the remarkable result  that ${\tau_v^+}_{min}$ depends on the  amplitude of noise and temperature 
 only through $\tau^L(+)$.
Interestingly,  the relationship between  ${\tau_v^+}_{min}$ and $\tau^L(+)$
is more complicated than what is suggested in Eq.(\ref{focusingcond}). Indeed 
in  Fig.~\ref{log}  the dashed line indicates a square root dependence, so that we have: 
\begin{equation}\label{regression}
 {\tau_v^+}_{min} \propto \sqrt{\tau^L(+)}
\end{equation}
We have no theoretical arguments suggesting the scaling law above, 
and more investigation is required.

\begin{figure}[h]
\spazio
 \centering
 \includegraphics[width=0.6 \textwidth]{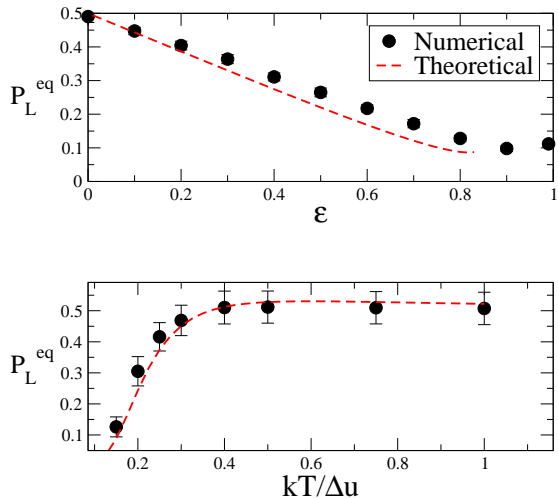}
  \caption{(Color online) 
Upper panel:  $P_L^{eq}$ at focusing (black dots)  is plotted {\it vs} $\epsilon$ for $D/\tau_v= 0.71$, $k_B T = 0.15 \Delta u$.
Lower panel: $P_L^{eq}$ at focusing (black dots)  is plotted {\it vs} $k_BT/ \Delta u$ for $D/\tau_v = 10 (k_BT)^2$, $ \epsilon= 0.8$.
In both panels the dashed lines represent the theoretical predictions obtained from the master equation, see text.
}
 \label{parameters}
\end{figure}

Up to now, we have only studied the probability distribution induced by the DN as a 
function of $\tau_v$. Now we intend to analyze the 
dependence of the focusing  on the other parameters, such as $\epsilon$ and $k_B T$. 
In Fig.~\ref{parameters} (upper panel) we show the left well  probability at the focusing, $P_L^{eq}$ {\it vs} 
the DN asymmetry parameter, $\epsilon$. We observed that
the focusing decreases as the asymmetry decreases ($\epsilon \rightarrow 0$). This can be explained by notecing that the distance between the two thermal transition times $\tau^L(+)$ and $\tau^R(-)$ decreases as the asymmetry decreases, therefore the second focusing condition in Eq.(\ref{focusingcond}) is not met.  
In Fig.~\ref{parameters} (lower  panel)   we show the left well  probability at the focusing, $P_L^{eq}$ {\it vs} the temperature. 
 Obviously the  temperature destroys the focusing effect.
 
From Fig.~\ref{minimo} we noticed  that  the master equations fail to fit the data when $\tau_v$ is smaller than the value at which a minimum is found. After that point, the equilibrium probability computed with the master equations reaches the limiting value, $P_L^{for}$, given in Eq.(\ref{PLfoc}), which is 
close to $P_L^{eq} $, evaluated at the focusing.
Thus we can use   Eq.(\ref{PLfoc}) to analytically estimate the value of the  focusing as a function of the parameters.
The result  is shown in both panels of Fig.~\ref{parameters} as dashed curves, which are in good agreement with our  numerical results,
 as long as the asymmetry, $\epsilon$, and the temperature, $k_B T$, are not too large. 
 Indeed in these cases the Kramers formula (\ref{Kramers}) is no longer applicable since the energy barriers are not well defined.

 The results of our investigations on a double well potential 
 can be summarized as follows:
\begin{itemize}
 \item a stochastic perturbing potential with zero time average can induce an unbalanced stationary probability distribution between the two wells. 
 
\item conditions  for the right-well focusing are: $\tau_v^+ \simeq \tau^L(+)$ and $\tau_v^- \ll  \tau^R(-)$, see Eq.(\ref{focusingcond}). Similarly for the left-well focusing.

\item At fixed DN intensity  the focusing can be improved by increasing the asymmetry $\epsilon$ or decreasing the temperature. 
Analytical estimation of the focusing intensity is given in Eq.(\ref{PLfoc}).

\item   The master equation approach completely fails  for fast DN perturbations (small $\tau_v$)  and is unable to describe
the  \textit{nonequilibrium kinetic focusing}. 

\end{itemize}


\section{Eight-well potential: a more realistic model for ion channel gating}
\begin{figure}[h]
\spazio
 \centering
 \includegraphics[width=0.5\textwidth]{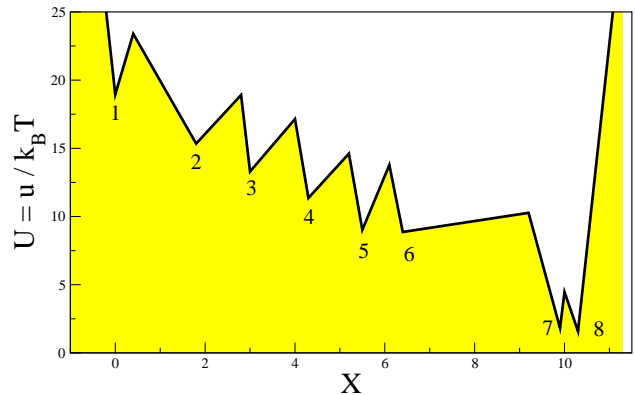}
 \caption{The potential $u$ used by Millonas and Chialvo in \cite{Millonas} as a model of the Shaker $K^+$ channel. For the definition of the rescaled potential $U(X)$, see Eq.~(\ref{rescaling}). In this Figure we assumed $T= 300K$.}
 \label{8well-pot}
\end{figure}

In the previous Section we showed  that it is possible to modify the \textit{stationary} probability distribution in a double-well potential, and particularly to increase the probability to find the system in one  well,
 by introducing a stochastic perturbation. Although a potential with only two states can be a useful starting point, 
 a multiwell potential is a more realistic model for ion channels. 
As an example, we consider, as in \cite{Millonas}, the Shaker $K^+$ channel, for which there exists a simple kinetic model, proposed by Bezanilla, Perozo and Stefani (BPS) \cite{Bezanilla}: in this model, there are eight states, or conformations, and the system can jump between them.
All the states from $1$ to $7$ correspond to  closed conformations, and  the state $8$ is the only one for which the ion channel is open (which means that
the ions can go through the channel).  
The allowed transition rates  for the Shaker $K^+$ ion channel are known from experimental measurements. 
 Since transitions occur only 
between nearest  neighboring states, this can be  represented
 by an eight-well potential landscape $u(x)$, as depicted in  Fig.~\ref{8well-pot}.
 The eight potential minima are the possible states of the channel,
 and the barriers heights are chosen so to reproduce the correct exponential transition rates.

In the following we present some preliminary discussion on the effect of stochastic noise on a multiwell potential landscape,
leaving a deeper analysis for future work.

The problem of focusing in this eight-well model has been already considered by 
Millonas and Chialvo \cite{Millonas}.
Based on a probabilistic treatment of the Langevin equation (\ref{LangMill}), Millonas and Chialvo  claimed that,
for very low temperature and large amplitude of the DN,
it is possible to induce \textit{stationary} probability distributions
 located in a chosen potential well: they called this effect \textit{nonequilibrium kinetic focusing}. 
These two conditions (low temperature and large DN) are not physiological, since the ion channel would be destroyed 
under these conditions.
For this reason, we will first investigate whether 
 a \textit{nonequilibrium kinetic focusing} 
can occur in an eight-well potential in physiological conditions (room temperature and moderate DN amplitude)
as it was considered in the double well case in the previous Section. 
We will  also compare the analytical predictions of Ref. \cite{Millonas} for low temperature and large DN intensity with our numerical simulations.

\subsection{Room temperature and low DN intensity}

Our starting point is the Langevin equation (\ref{LangMill}), and the analysis of the results
 will concern the equilibrium probabilities $P^{eq}_i$ in each well, defined as in the previous Sections.

\begin{figure}[h!]
 \spazio
 \centering
 \includegraphics[width=0.54 \textwidth]{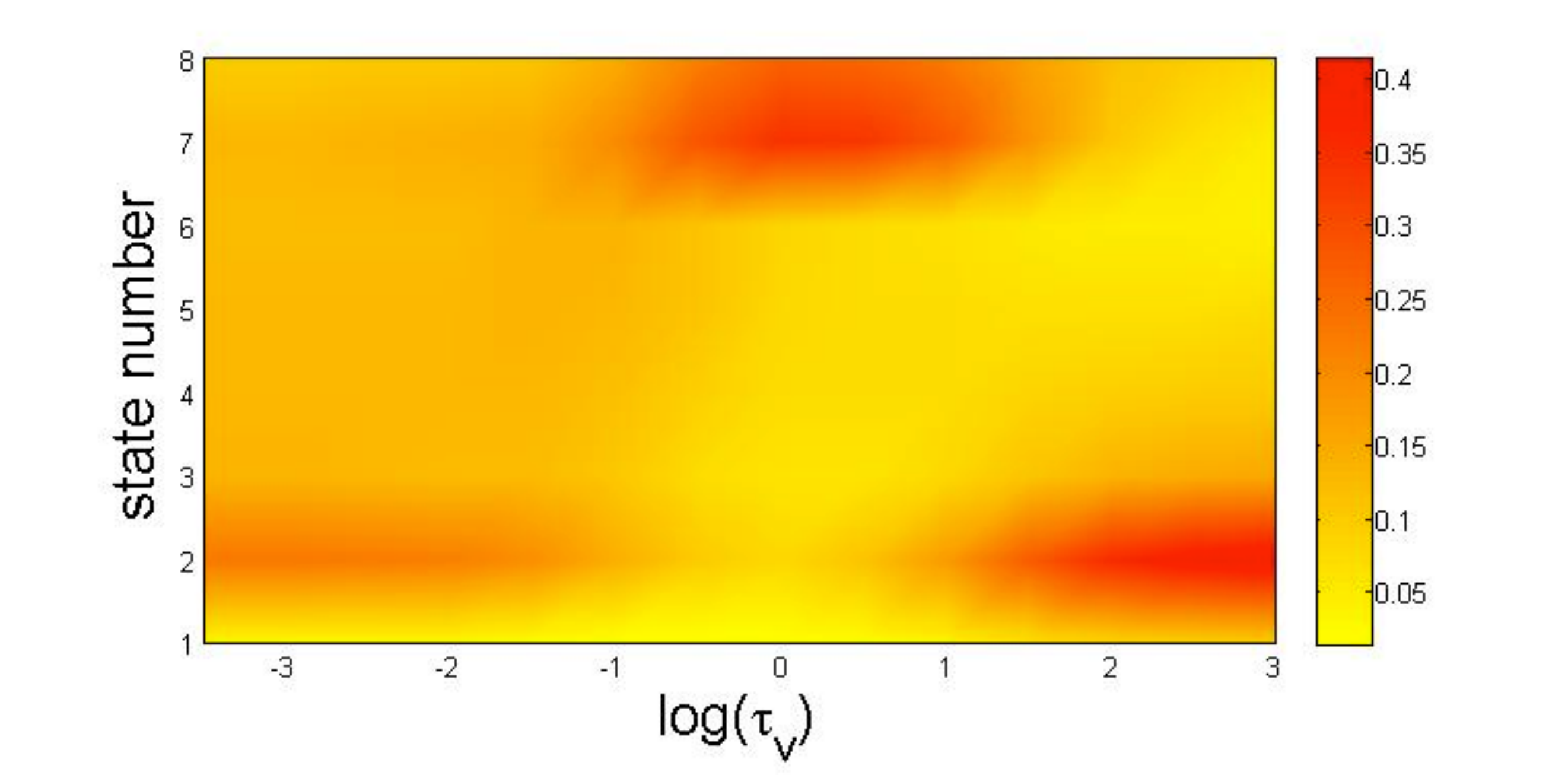}
 \caption{(Color online) Density plot of the equilibrium probabilities $P_i^{eq}$ in the i-th  well 
versus the DN correlation  time $\tau_v$. Darker color intensities stand for higher probabilities.
Data are: $D/\tau_v=1$, 
 $\epsilon=0.8$, $V_0=\langle V_{ext}\rangle=(-45\textrm{ mV})/k_BT$, and $T=300\textrm{ K}$. 
 }
 \label{8well-eps8}
\end{figure}

In Fig.~\ref{8well-eps8} we report our simulations concerning the equilibrium probability distribution $P^{eq}_i$ as a function of $\tau_v$
for each well, at fixed asymmetry 
$\epsilon$, intensity $D/\tau_v$ and mean external potential $V_0$.
Here we can see that
some kind of focusing occurs  only close to the potential  edges (states $1,2$ or $7,8$), 
while the equilibrium probabilities of the central wells are almost independent of $\tau_v$.

From  the same figure it is also possible to recover the usual observations made in the double-well case about the limits of very small and very large $\tau_v$. 
We checked that when $\tau_v \rightarrow 0$, the DN averages to zero, so that  the probability distribution 
the thermal Gibbs distribution for $V_{ext}=V_0$. 
On the other hand, for $\tau_v \gg 1$ the probabilities can be evaluated, as in Eq.(\ref{limitedx}), by
\begin{equation}\label{pmean}
P_i^{mean} = \frac{\tau_v^+}{(\tau_v^+ + \tau_v^-)}P_i
(+)+\frac{\tau_v^-}{(\tau_v^+ + \tau_v^-)}P_i(-),
\end{equation}
where the probabilities $P_i(\pm)$ are the thermal Gibbs  probabilities for each well in the $(+)$ and $(-)$ 
configuration.

The non trivial focusing in the intermediate $\tau_v$ regime, resembles the one found in the double well case.
Indeed the conditions given in Eq.(\ref{focusingcond}) seem to work also in the multi-well case if we consider the edge wells $1,2$ as a left well; the edge wells
$7,8$ as the right well and the central wells as an effective barrier between them.  
Nevertheless a more detailed analysis is necessary to understand how to relate the focusing in the double well with that
 in the multi-well case.

We also considered many other situations, with different values of the DN asymmetry parameter $\epsilon$ and of the DN intensity $D/\tau_v$: in all cases, 
we could only observe some probability increase in the wells on the edges, while, to the best of our effort,
we were not able to obtain focusing in the central wells. 
Such a result would be desirable since it would allow a complete control of the ion channel dynamics
under a stochastic perturbations.

\subsection{Low temperature and large DN intensity}

An analytical solution (Eq.$(7)$ in  Ref.~\cite{Millonas}) for the stationary probability distribution induced by a DN,
were derived under the following assumptions:
\begin{itemize}
 \item the amplitude of the driving, must be larger than  the thermal fluctuations,
\item and larger than the maximal potential jump, $\sqrt{D/\tau_v}>\sup|U'|$;
\item the temperature should be  very small (that is $T\rightarrow 0$).
\end{itemize}

 In the above Section, we did not take these conditions into account, because we considered them not physiologically feasible,
 nevertheless in this Section we take this regime into consideration in order to understand the range of validity 
 of the analytical prediction given in Ref.~\cite{Millonas}).

\begin{figure}
\spazio
 \centering
 \includegraphics[width=0.54 \textwidth]{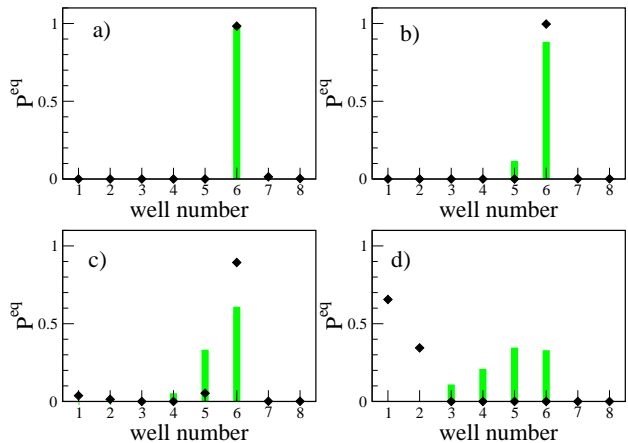}
 \caption{(Color online) Probability distributions at different temperature
 computed with the Langevin equation (green bars) and theoretically (black diamonds) 
 with Eq.$(7)$ in  Ref.~\cite{Millonas}.
The  parameters are: $D/\tau_v=10^5$, 
$\epsilon = 0.6$, $\tau_v=10^{-4}$, $v_0=-142 \textrm{ mV}/k_BT$.
Temperature is : $T= 20 K$ (a), $T= 25 K$ (b), $T= 30 K$ (c), $T= 40 K$ (d).
}
 \label{controesempio2}
\end{figure}

 In Fig.~\ref{controesempio2}
  we compare  our numerical results (hystogram) with the theoretical results (dots) given in  Ref.~\cite{Millonas}
 for different temperatures while fixing all the other parameters. 
 As we can see, while at low temperature (panel a) and b)) there is a good agreement,  on increasing the temperature (panel c) and d))
 the theoretical 
distribution  starts to fail  even for temperature well below room temperature ($ T= 40 \textrm{ K}$). 
Interestingly we notice a good focusing in the central wells: in panel (c)  Langevin equation gives approximately
90 \% of probability in the wells 5 and 6), even when the theoretical equation of Ref.~\cite{Millonas} does not work. 
 The presence of some focusing
in this case seems to indicate that the conditions given in  Ref.~\cite{Millonas}, 
could be too restrictive 
in order to have  focusing.

\section{Conclusions}

In this work we considered two different kinetic models of ion channels  subjected to dichotomous noise perturbation,
with zero time average.  The first  is a simple model consisting of two wells separated by an energy barrier, while the second
is the physiologically relevant eight-well model  proposed by Bezanilla et al. \cite{Bezanilla}. 
In both cases an highly non trivial non equilibrium focusing has been found under physiological conditions (room temperature
and moderate external perturbation intensity), which cannot be described
by a master equation approach, but the analysis of the full Langevin equation turns out to be  necessary. 
Conditions for the focusing and dependence on the external perturbation parameters have been given for the 
double well case.  In the eight-well case more analysis is needed in order to address this question. 

We also analyzed the results of  Millonas and Chialvo in \cite{Millonas}, showing that the
\textit{nonequilibrium kinetic focusing} does not survive in physiological conditions.

In perspective our analysis leaves still  open the question of whether it is possible to focus an ion channel in any
well, thus completely controlling its dynamics, under physiological conditions. Nevertheless we have shown both that
focusing is possible in physiological conditions and also we gave a preliminar evidence that focusing
in the central wells of a multi-well potential is also possible.


\end{document}